\documentclass[lettersize,journal]{IEEEtran}

\IEEEoverridecommandlockouts

\usepackage{amsfonts}
\usepackage[dvips]{graphicx}
\usepackage{times}
\usepackage{cite}
\usepackage{amsmath}
\usepackage{cases}
\usepackage{array}
\usepackage{dsfont}
\usepackage{amssymb}

\usepackage{makecell}
\usepackage{stfloats}
\usepackage{booktabs}
\usepackage{graphicx}
\usepackage{subfigure}
\usepackage{pdfpages}
\usepackage{footnote}
\usepackage{soul}
\usepackage{array}
\usepackage{multirow}
\usepackage{bm}
\usepackage{empheq}
\usepackage{amsthm}
\usepackage{algorithm}
\usepackage{algorithmicx}
\usepackage{algpseudocode}
\usepackage{color}
\usepackage{diagbox}
\usepackage{caption}
\theoremstyle{plain}

\theoremstyle{definition}

\newtheorem{remark}{Remark}

\usepackage[colorlinks,bookmarksopen,bookmarksnumbered,citecolor=blue, linkcolor=blue, urlcolor=blue]{hyperref}

\let\oldReturn\Return
\renewcommand{\Return}{\State\oldReturn}

\usepackage{textcomp}
\usepackage{xcolor}
\ifCLASSINFOpdf
\else
\fi

\begin{document}
	\title{CTD-Diff: Cooperative Time-Division Diffusion for Multi-User Semantic Communication Systems}
	\vspace{-4pt}
	
	\vspace{-25pt}\author{\IEEEauthorblockN{$\text{Chengyang Liang}$, $\text{Dong Li}, ~\IEEEmembership{Senior Member,~IEEE}$}
		\vspace{-22pt}
		\thanks{Chengyang Liang and Dong Li are with the School of Computer Science and Engineering, Macau University of Science and Technology, Macau, China (e-mail: 3240006992@student.must.edu.mo; dli@must.edu.mo).}}
	\vspace{-20pt}

\maketitle
\begin{abstract}
	
	Semantic communication (SemCom) has emerged as a transformative paradigm for efficient information transmission by emphasizing the exchange of task-relevant meaning rather than raw data. While diffusion-based SemCom models have demonstrated remarkable generative capabilities, existing studies predominantly focus on point-to-point links, overlooking the potential of multi-user (MU) cooperation in MU wireless environments. To address this limitation, we propose a Cooperative Time-Division Diffusion (CTD-Diff) framework. Unlike traditional approaches that view channel noise solely as a detriment, our framework innovatively integrates the noisy wireless transmission process directly into the forward diffusion chain. Specifically, we design a multi-user cooperation mechanism based on Time-Division Multiple Access (TDMA), where idle users overhearing the active transmitter act as semantic collaborators. To maximize the signal fidelity, the receiver employs direct signal aggregation to fuse the direct signal with cooperative copies. This aggregated noisy semantic representation serves as the condition for the reverse diffusion process, allowing the receiver to reconstruct high-fidelity data by mitigating the cumulative channel distortions. By effectively converting physical channel noise into diffusion noise, the proposed method significantly enhances the transmission reliability. Extensive experiments demonstrate that CTD-Diff outperforms various baselines regarding the reconstruction accuracy and the perceptual quality, particularly under challenging low signal-to-noise ratio (SNR) conditions.
	
\end{abstract}


\IEEEpeerreviewmaketitle

\vspace{-4pt}
\section{Introduction} 
\label{sec:Introduction}

\IEEEPARstart{T}{he} rapid proliferation of data-intensive and intelligent applications, including autonomous driving, virtual reality, and multimodal human–machine interaction, has imposed unprecedented demands on wireless communication systems \cite{Gu, Rethinking, FromData}. Traditional communication paradigms, grounded in Shannon’s bit-level information theory, primarily emphasize the reliable transmission of symbol sequences without consideration of the semantic content or task relevance of the transmitted data. However, bit-accurate transmission is increasingly inefficient and unnecessary for intelligent tasks, where the primary objective is to convey semantic meaning rather than exact signal realizations. In response to this paradigm shift, semantic communication (SemCom) has emerged as a transformative framework that integrates communication theory, machine learning (ML), and signal processing to facilitate end-to-end transmission of task-relevant information \cite{GenAIsemcomLiang}. SemCom aims to minimize the task-level distortion, such as semantic reconstruction error or perceptual loss, by jointly optimizing the source encoder and channel encoder through deep learning (DL) techniques, rather than focusing on the conventional bit error rate (BER). This approach has demonstrated significant potential in reducing the bandwidth consumption, enhancing the robustness under low signal-to-noise ratio (SNR) conditions, and achieving human-level understanding in image, speech, and text transmission scenarios \cite{Video, NLJSCNSC, SpeechReg, SCcarrier, CommunicateLess}.

Recent advances in DL have further propelled the evolution of SemCom by enabling powerful representation learning across complex and high-dimensional data modalities. Through large-scale training and end-to-end optimization, deep neural networks can learn hierarchical semantic features that capture task-relevant information beyond conventional compression and channel coding \cite{WITT, DeepJSCC, DJSCCforSemCom}. Building on this foundation, generative artificial intelligence (GenAI) models have emerged as a new frontier in semantic representation and synthesis, allowing communication systems to generate, reconstruct, or imagine semantic contents even under limited or corrupted channel observations \cite{GANBasedSemCom, GAIsemcom, TcomLiang, GAIATA}. In particular, diffusion models, a category of likelihood-based generative models, have recently achieved state-of-the-art (SOTA) performance in image generation tasks. In a standard denoising diffusion probabilistic model, a forward noising process  is defined as $x_t=\sqrt{\bar{\alpha}_t}x_0+\sqrt{1-\bar{\alpha}_t}\epsilon$ with $\epsilon\sim\mathcal{N}(0,I)$, and a neural network $\epsilon_\theta(x_t,t)$ is then trained to predict the noise component. During the reverse process, the model iteratively denoises $x_{t}$ to recover the original data $x_{0}$. Recently, diffusion-based generative approaches have been increasingly adopted in SemCom research. For instance, the study in \cite{CDDM} incorporates a diffusion module as a receiver-side denoiser specifically designed to mitigate channel effects, while \cite{LightweightDM} utilizes proposed post-training quantization technique to reduce the computational load and memory footprint required for generating images using the diffusion model. Additionally, \cite{LatentDiffSemCom} introduces a latent-space diffusion approach to address potential outliers in source data. These works validate the potential of diffusion models to effectively bridge channel noise and semantic reconstruction. Nevertheless, existing works predominantly focus on single-user scenarios or latent-space mappings and do not explicitly tackle the challenges associated with multi-user (MU) interference management.

In practical wireless communication systems, MU communication is prevalent rather than exceptional, and achieving reliable semantic transmission necessitates approaches beyond treating users as isolated links \cite{MM}. MU scenarios inherently involve inter-user coupling through shared spectral resources, heterogeneous semantic priorities, and variable channel conditions, thereby requiring cross-user coordination instead of merely mitigating interference \cite{MUWireless}. Recent studies have begun to investigate these MU semantic interactions. For instance, \cite{KnowledgeDistillation} examines users sharing a common channel and employs a knowledge-distillation strategy that enables a compact student model to generalize robustly in MU coexistence scenarios, thereby demonstrating the advantages of shared semantic representations across users. Similarly, \cite{DMCE} utilizes diffusion models to restore and fuse semantic features from multiple users, illustrating that generative models can effectively integrate multi-user information rather than processing each user independently \cite{InterferenceSuppressed}. Furthermore, \cite{Semanticimportance} proposes a multi-input multi-output (MIMO) system that incorporates singular value decomposition (SVD)-based subchannel allocation alongside semantic importance modeling across users, highlighting the benefits of coordinating physical-layer resources with semantic priorities. Collectively, these studies suggest that MU SemCom requires joint modeling, cross-user feature integration, and co-design of resources and semantics \cite{tradeoffsc, BeamformingDesign}. Nevertheless, existing MU semantic frameworks often neglect cooperative overhearing opportunities, or fail to fully integrate physical-layer combining with generative semantic reconstruction.

Despite considerable progress in SemCom research, several fundamental challenges remain unresolved. First, the majority of existing diffusion-based frameworks primarily concentrate on point-to-point links \cite{DiffusionSC, JSCNA, LightweightDM, LatentDiffSemCom}, often neglecting the potential benefits of MU cooperation in MU scenarios. Although recent studies on MU SemCom have addressed resource allocation, they generally treat the MU environment as a source of contention to be managed rather than as an opportunity to be exploited. Consequently, the intrinsic temporal structure of time-division multiple access (TDMA) remains underutilized. The idle time slots of non-transmitting users are wasted instead of being utilized for cooperative relaying to mitigate the channel fading. Second, current approaches generally treat the physical channel noise and the model's diffusion noise as uncoupled entities \cite{CDDM}. They lack a unified theoretical framework that models the accumulation of noise from both direct and relayed transmission paths within the forward diffusion process \cite{LightweightDM}. This separation impedes the receiver's capacity to optimally fuse semantic information copies and coherently adapt the generative denoising process to address the compounded distortions introduced by the channel fading from multiple users.

To bridge these gaps, this paper proposes a Cooperative Time-Division Diffusion (CTD-Diff) framework that extends diffusion-based SemCom to MU wireless environments. In contrast to conventional approaches that consider channel noise and generative noise as separate entities, CTD-Diff integrates the physical wireless transmission, including fading and cooperative relaying, directly into the forward diffusion process. Existing studies have neglected this unified perspective, except for an initial exploration in our previous work \cite{JSCNA}. Nevertheless, the critical issue of forward noise combining within cooperative scenarios was not addressed therein and has yet to be investigated. Specifically, a cooperative TDMA mechanism is designed, wherein idle users function as semantic collaborators. The receiver utilizes direct signal aggregation to optimally integrate direct and forwarded semantic information copies. This fused information, which contains an aggregated noise component, serves as the starting point for the reverse diffusion process, enabling the denoiser to reconstruct high-fidelity data by jointly mitigating channel distortions and semantic compression artifacts. Furthermore, a conditional diffusion network guided by semantic embeddings ensures content consistency throughout the generation process. Collectively, these design elements establish a unified generative-semantic architecture that transforms physical channel noise from a detrimental factor into a manageable probabilistic component, thereby advancing reliable and scalable communication for next-generation 6G networks.

The contributions of this paper are summarized as follows.

\begin{itemize}
	\item \textbf{Unified Framework Bridging Cooperative Communication and Diffusion Generation:} We propose CTD-Diff, a unified framework that integrates multi-user cooperative transmission with diffusion-based semantic reconstruction. In contrast to conventional approaches that consider channel noise as an external distortion, the proposed method embeds the aggregation of channel fading and noise into the forward diffusion process. This design enables the receiver to interpret the noisy aggregated signal as a valid intermediate diffusion state, allowing the denoising network to jointly mitigate channel impairments and semantic distortions within a unified probabilistic framework. Furthermore, this formulation establishes a principled connection between physical-layer stochasticity and generative modeling, providing a novel perspective for the design of semantic communication systems.
	\item \textbf{Cooperative TDMA Mechanism:} We propose a TDMA-based cooperative transmission protocol in which idle users opportunistically serve as semantic relays. By implementing transmitter-side pre-equalization for both direct and relayed links, the received semantic signals are inherently aligned at the base station, facilitating low-complexity direct aggregation. This approach enhances the effective SNR and offers a superior initialization for subsequent diffusion-based reconstruction, thereby establishing a synergy between physical-layer cooperation and semantic recovery. Furthermore, the proposed design redistributes computational complexity from the receiver to distributed users, rendering it suitable for practical deployment scenarios.
	\item \textbf{Semantic-Conditioned Diffusion Reconstruction:} To improve robustness in the presence of severe channel degradation, we propose a semantic-conditioned diffusion model. High-level semantic embeddings, derived from the received signal, are incorporated into the denoising network through cross-attention mechanisms, thereby directing the reconstruction process toward the appropriate semantic manifold. This approach effectively mitigates semantic ambiguity and prevents the generation of semantically inconsistent outputs under low-SNR conditions. Moreover, the conditional formulation constrains the solution space of the generative process, enhancing both reconstruction stability and convergence performance.
	\item \textbf{Hybrid-Noise Training Strategy:} We propose a hybrid noise training strategy that integrates both synthetic diffusion noise and realistic wireless channel distortions. By training the denoising network on this composite noise distribution, the model learns a generalized restoration capability that effectively bridges the gap between idealized diffusion assumptions and practical channel conditions. This approach substantially enhances robustness and generalization across varying SNR levels and multi-user scenarios. Furthermore, the proposed training scheme enhances the model’s adaptability to diverse channel statistics without necessitating retraining.
\end{itemize}

The rest of this paper is organized as follows. Section \ref{sec:system model} introduces the system model and problem formulation. Section \ref{sec:architecture} presents the proposed CTD-Diff framework, including the Conditional Diffusion Reconstruction and Hybrid-Noise Training Strategy. Section \ref{sec:Experiments} provides numerical experiments and ablation studies demonstrating the effectiveness of the proposed approach. Finally, Section \ref{sec:conclusion} concludes the paper.

\vspace{-8pt}
\section{System Model}
\label{sec:system model}
\vspace{-4pt}

\begin{figure*}[t]
	\centering
	\includegraphics[scale=0.72]{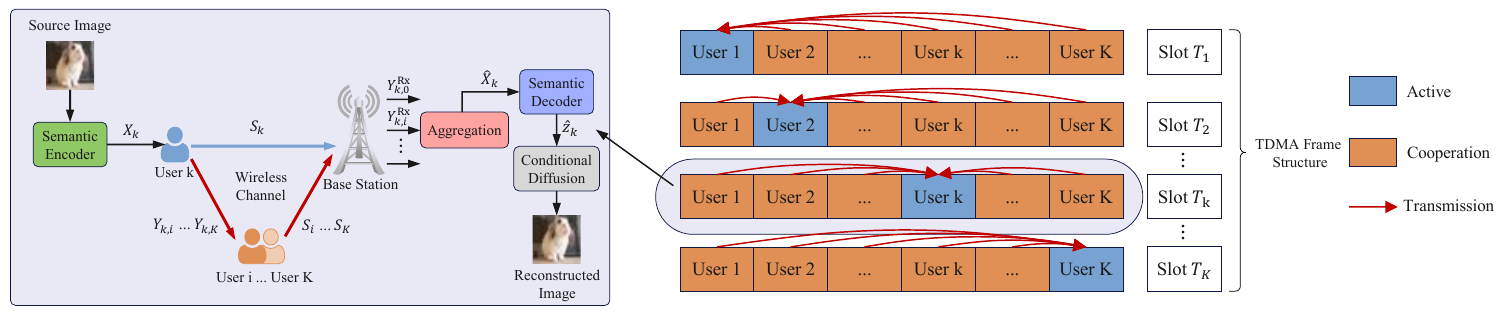}
	\caption{Illustration of the proposed semantic communication system.}
	\label{fig:system}
	\vspace{-10pt}
\end{figure*}

This section delineates the comprehensive system model of the proposed CTD-Diff framework. As illustrated in Fig. \ref{fig:system}, the framework unifies semantic encoding, cooperative wireless transmission, and diffusion-based reconstruction into a holistic pipeline. Specifically, we detail the transmitter-side pre-equalization, the cooperative semantic information aggregation at the receiver, and how the aggregated channel noise is strategically integrated into the semantic diffusion process.

\subsection{Transmitter Design}
\label{sec:transmitter}

Consider the MU communication system consisting of $K$ users, indexed by $\mathcal{U} = \{1, 2, \ldots, K\}$, and a central BS. Each user $k$ aims to transmit a high-dimensional image $I_k \in\mathbb{R}^{C\times H\times W}$, where $H$ and $W$ are the image height and width, and $C$ is the number of color channels. Prior to transmission, a semantic encoder $f_{\text{enc}}(\cdot)$ compresses $I_k$ into a compact semantic representation $X_k = f_{\text{enc}}(I_k)$. The system operates under a cooperative TDMA protocol. When user $k$ serves as the active source node, the remaining idle users $i \in \mathcal{U}\setminus\{k\}$ act as semantic collaborators to assist in the transmission. To enable low-complexity semantic aggregation at the receiver, a channel inversion pre-equalization strategy is employed at both the source and collaborating user nodes.

\subsubsection{Source Transmission (User $k$)}
During its designated transmission slot, user $k$ transmits the semantic feature $X_k$. To counteract the fading effects of the direct link to the BS ($h_{k,0}$), user $k$ employs pre-equalization by scaling the semantic information with the inverse of the channel coefficient. Consequently, the transmitted signal $S_k$ is expressed as $S_k = \frac{1}{h_{k,0}} X_k,$ where $h_{k,0}$ represents the channel coefficient from user $k$ to the BS. This approach guarantees that the direct semantic component received at the BS corresponds precisely to the original semantic feature $X_k$.

\subsubsection{Cooperative Forwarding (User $i$)}
Simultaneously, the broadcast semantic feature $S_k$ is overheard by idle users. The semantic information received by an idle user $i$ is given by
\begin{equation}
	Y_{k,i} = h_{k,i} S_k + N_{i} = \frac{h_{k,i}}{h_{k,0}} X_k + N_{i},
\end{equation}

\noindent where $h_{k,i}$ denotes the channel from user $k$ to user $i$, and $N_{i}$ represents the receiver noise.

Upon receiving $Y_{k,i}$, the user $i$ executes semantic forwarding. To guarantee that the forwarded semantic information arrives at the BS aligned with $X_k$, the user undertakes a two-step normalization process: first, compensating for the first-hop channel $h_{k,i}$ and second, pre-equalizing for the second-hop channel $h_{i,0}$. The signal transmitted by user $i$ is thus expressed as

\begin{equation}
	S_{i} = \frac{1}{h_{i,0}} \left( \frac{h_{k,0}}{h_{k,i}} Y_{k,i} \right) = \frac{1}{h_{i,0}} X_k + \frac{h_{k,0}}{h_{i,0} h_{k,i}} N_{i}.
\end{equation}

\noindent By dividing the channel coefficients at the transmitter, the computational burden of channel equalization is transferred from the BS to the distributed users, thereby facilitating a simplified aggregation mechanism at the receiver.

\subsection{Receiver Design}
\label{sec:receiver_combining}

The BS obtains semantic information from both the direct link associated with user $k$ and additional links from other users $i$. Due to the pre-equalization applied at the transmitters, the fading effects on the semantic feature $X_k$ are effectively mitigated prior to reception. The semantic feature received via the direct link is expressed as follows

\begin{equation}
	Y_{k,0}^{\text{Rx}} = h_{k,0} S_k + N_0 = X_k + N_0.
\end{equation}

\noindent Similarly, the semantic feature received from user $i$ is given by
\begin{equation}
	Y_{k,i}^{\text{Rx}} = h_{i,0} S_{i} + N_0^{(i)} = X_k + \underbrace{\left( \frac{h_{k,0}}{h_{k,i}} N_{i} + N_0^{(i)} \right)}_{\tilde{N}_i},
\end{equation}

\noindent where $N_0$ and $N_0^{(i)}$ denote the noise at the BS corresponding to the respective links.

Since all received copies $\{Y_{k,0}^{\text{Rx}}, Y_{k,1}^{\text{Rx}}, \ldots\}$ contain the identical semantic component $X_k$, the BS utilizes a direct signal aggregation method by summing the received copies
\begin{equation}
	Y_k^{\text{agg}} = Y_{k,0}^{\text{Rx}} + \sum_{i \in \mathcal{U}\setminus\{k\}} Y_{k,i}^{\text{Rx}} = (R+1) X_k + N_{\text{agg}},
\end{equation}
\noindent where $R$ denotes the number of idle cooperative users, and $N_{\text{agg}}$ represents the aggregated noise term. After normalization by the $(R+1)$, the resulting aggregated observation is derived as $\hat{X}_k = X_k + \tilde{N}_k,$ where $\tilde{N}_k = N_{\text{agg}}/(R+1)$ represents the effective combined channel noise.

\begin{remark}
	Traditional communication systems typically regard noise aggregation as harmful. In contrast, within our CTD-Diff framework, this aggregated noise fulfills a beneficial role. Through channel pre-equalization, the semantic feature $X_k$ is constructively enhanced, while the noise components accumulate to form a complex Gaussian distribution. Importantly, we interpret this aggregated channel noise $\tilde{N}_k$ not as interference to be eliminated, but rather as the initial noise perturbation that drives the subsequent diffusion generation process.
\end{remark}

\vspace{-14pt}
\subsection{Diffusion-based Semantic Reconstruction}
\label{sec:diffusion}

The aggregated semantic feature $\hat{X}_k$ comprises the original semantic information $X_k$ that has been degraded by physical channel noise $\tilde{N}_k$. This noisy observation is conceptualized as an intermediate state within a diffusion process, thereby linking the physical channel to the generative model.

\subsubsection{Forward Process as Channel Distortion}
Standard diffusion models characterize a forward process that incrementally introduces Gaussian noise to the clean data. In our framework, we map the physical transmission to this process. The aggregated observation $\hat{X}_k$ is regarded as a noisy state $X_{k, t_{\text{ch}}}$ at a particular timestep $t_{\text{ch}}$ within the diffusion chain
\begin{equation}
	\underbrace{\hat{X}_k}_{\text{Observation}} \equiv X_{k, t_{\text{ch}}} \approx \sqrt{\bar{\alpha}_{t_{\text{ch}}}} X_k + \sqrt{1-\bar{\alpha}_{t_{\text{ch}}}} \epsilon,
\end{equation}
\noindent where the effective channel noise $\tilde{N}_k$ inherently resembles the Gaussian noise $\epsilon$ utilized in the diffusion model. This correspondence obviates the necessity for artificial noise injection at the onset of inference, thereby transforming the detrimental physical channel noise into a controllable probabilistic element.

\subsubsection{Reverse Denoising Process}
The reconstruction is conducted using a conditional denoising network, denoted as $\epsilon_\theta(X_{k,t}, t)$, which is trained to reverse the diffusion process. In contrast to conventional receivers that necessitate distinct channel estimation and equalization components, the proposed denoiser simultaneously performs channel noise mitigation and semantic reconstruction. The training objective involves minimizing the noise prediction error:
\begin{equation}
	\mathcal{L}_{\mathrm{diff}} = \mathbb{E}_{X_k, t, \epsilon \sim \mathcal{N}(0,I)} \left[ \| \epsilon - \epsilon_\theta(X_{k,t}, t) \|_2^2 \right],
\end{equation}
\noindent where $X_{k,t}$ denotes the noisy state obtained from the clean feature $X_k$ at timestep $t$. During inference, the BS begins with the channel-corrupted observation $\hat{X}_k$ (treated as starting state $X_{k, t_{\text{ch}}}$) and iteratively performs the reverse sampling procedure to reconstruct the high-fidelity semantic feature $X_k$:
\begin{equation}
	X_{k,t-1} = \frac{1}{\sqrt{\alpha_t}} \left( X_{k,t} - \frac{1-\alpha_t}{\sqrt{1-\bar{\alpha}_t}} \epsilon_\theta(X_{k,t}, t) \right) + \sigma_t \mathbf{z},
\end{equation}
\noindent where $\mathbf{z} \sim \mathcal{N}(0, I)$ is the Gaussian noise. Through this unified architecture, CTD-Diff converts the integration of physical layer noise into a generative sampling process, thereby enabling robust SemCom without the need for complex channel processing at the receiver.

\vspace{-6pt}
\section{CTD-Diff Framework and Cooperative Mechanism}
\label{sec:architecture}

This section delineates the fundamental technical components of the CTD-Diff framework alongside its collaborative transmission mechanism. While the system model establishes the foundational workflow for MU TDMA transmission and signal aggregation, several critical challenges persist: the deep integration of the physical-layer cooperative mechanism with the semantic-layer diffusion generative model; the design of conditional denoising networks that effectively exploit semantic priors; and the reconciliation of the statistical discrepancies between theoretical diffusion processes and real-world wireless channel distortions via hybrid noise training strategies.

Initially, this section provides an overview of how CTD-Diff integrates cooperative communication within a probabilistic generative framework. Subsequently, it elaborates on the design rationale and implementation details of the conditional diffusion reconstruction network. Finally, it discusses how the hybrid noise training strategy improves the model's generalization capabilities under practical channel conditions.

\vspace{-6pt}
\subsection{CTD-Diff Framework}
\label{sec:CTD-Diff Framework}

Traditional SemCom systems generally consider the physical layer channel as an external source of disturbance that is independent of the semantic encoding and decoding processes. At the receiver, channel impairments are typically addressed through channel equalization or error-correcting codes prior to semantic decoding. This decoupled approach neglects the potential statistical correlation between channel randomness and the semantic generation process, resulting in significant performance degradation under conditions of low SNR or severe fading. In contrast, the principal innovation of the CTD-Diff framework is the integration of multi-user cooperative transmission, channel distortion, and diffusion generation models into a unified end-to-end stochastic Markov process. Specifically, the aggregated channel noise is no longer considered interference to be eliminated. Rather, it is treated as the initial noise input effectively advancing the data to an intermediate state of the diffusion forward process. This methodology enables the reverse denoising process to simultaneously address both artificial diffusion noise and genuine physical channel distortion.

This section examines the semantic information flow within the system model. The semantic information $X_k$ transmitted by user $k$ undergoes pre-equalization, multipath fading, and cooperative relaying. Subsequently, the BS acquires the aggregated observation $\hat{X}_{k}$ through direct signal aggregation, as derived in Section \ref{sec:receiver_combining}. In the conventional diffusion model, the forward process begins with clean data $X_k$ and progressively adds Gaussian noise until $X_{T}\sim\mathcal{N}(0,I)$. In CTD-Diff, we map the physical channel distortion to the initial phase of this process. Specifically, the received observation $\hat{X}_{k}$ is modeled as the state $X_{k, t_{\text{ch}}}$ at a specific timestep $t_{\text{ch}}$. Consequently, the diffusion chain constitutes a hybrid stochastic process that incorporates both intrinsic randomness arising from the physical channel and artificially introduced diffusion noise. This hybrid characteristic allows the forward process distribution at any step $t$ to be formulated relative to the clean semantics $X_k$
\begin{equation}
	q(X_{k,t}\mid X_{k})=\mathcal{N}\left(\sqrt{\bar{\alpha}_t}X_{k},(1-\bar{\alpha}_t)I\right).
\end{equation}
\noindent At the specific step $t=t_{\text{ch}}$, since $\hat{X}_{k}$ is a noisy observation, the variance matches the channel conditions: $\mathrm{Var}[X_{k,t_{\text{ch}}}]=(1-\bar{\alpha}_{t_{\text{ch}}})I \approx \Sigma_{\tilde{N}_k}$, where $\Sigma_{\tilde{N}_k}$ represents the covariance matrix of the effective channel noise. This composite structure suggests that the initial stages of the diffusion process are predominantly influenced by channel distortion. As $t$ increases beyond $t_{\text{ch}}$, $\bar{\alpha}_{t}$ approaches zero, allowing the artificial diffusion noise to progressively dominate. Ultimately, $X_{k,T}$ converges to a standard Gaussian distribution.

The objective of the denoising process is to recover the clean semantic feature $X_k$ starting from the channel-corrupted observation $\hat{X}_{k}$. Since the forward process incorporates two sources of noise, the denoising network in the backward process must be trained to simultaneously identify and eliminate both types of perturbations. According to the theory of denoising diffusion probabilistic models, the transition kernel in the backward process can be approximated by a Gaussian distribution
\begin{equation}
	p_\theta(X_{k,t-1}\mid X_{k,t})=\mathcal{N}\left(\mu_\theta(X_{k,t},t),\sigma_t^2I\right),
\end{equation}
\noindent where the mean function is parameterized by the noise prediction network $\epsilon_{\theta}(\cdot)$:
\begin{equation}
	\mu_\theta(X_{k,t},t)=\frac{1}{\sqrt{1-\beta_t}}\left(X_{k,t}-\frac{\beta_t}{\sqrt{1-\bar{\alpha}_t}}\epsilon_\theta(X_{k,t},t)\right).
\end{equation}
\noindent The objective function for training this network is:
\begin{equation}
	\mathcal{L}_{\mathrm{CTD-Diff}}=\mathbb{E}_{X_k, t, \epsilon}\left[\|\epsilon-\epsilon_\theta(X_{k,t},t)\|_2^2\right],
\end{equation}
\noindent where $\epsilon$ denotes the standard Gaussian noise component inherent in the noisy state $X_{k,t}$. In our hybrid training strategy, for steps $t \le t_{\text{ch}}$, this $\epsilon$ corresponds to the normalized physical channel noise. By minimizing this noise prediction error, the network implicitly acquires the statistical characteristics of channel distortion. From the standpoint of probabilistic inference, the training procedure corresponds to maximizing the likelihood of the clean data $X_k$ under the reverse Markov chain
\begin{equation}
	\begin{aligned}
		\mathbb{E}_q[\log p_\theta(X_{k})]
		&\ge \mathbb{E}_q \left[ \log p(X_{k,T}) \right.\\
		&\qquad\left.+ \sum_{t=1}^T \log p_\theta(X_{k,t-1}\mid X_{k,t})\right].
	\end{aligned}
\end{equation}
\noindent Under the regularity assumption, the gradient of the variational lower bound is proportional to the noise prediction error. Consequently, minimizing $\mathcal{L}_{\mathrm{CTD-Diff}}$ is equivalent to learning the optimal posterior estimator for $X_k$.

The cooperative transmission mechanism is fundamental to the entire framework. Multiple spatially distributed relays deliver aligned semantic copies. Instead of employing complex weighted combining strategies, the receiver utilizes a direct signal aggregation mechanism. This approach coherently reinforces the semantic features while aggregating the independent noise components from multiple paths. This fused observation serves as the robust starting point $X_{k, t_{\text{ch}}}$ for the reverse denoising process, effectively bridging physical transmission and semantic generation.

Consequently, the CTD-Diff framework facilitates cross-layer synergy between physical-layer collaboration and semantic-layer generation. Collaborative transmission enhances the initial conditions of the diffusion model input, while the diffusion model uniformly addresses residual distortion following collaborative merging through a probabilistic generation framework. These two components mutually reinforce one another, establishing an end-to-end optimized system.

\vspace{-8pt}
\subsection{Conditional Diffusion Reconstruction}
\label{sec:Conditional Diffusion}

The conventional diffusion model depends exclusively on the time step $t$ and the noise sample $X_{k,t}$ during the generation process, lacking the capability to incorporate external semantic constraints to guide reconstruction. This limitation is particularly pronounced in MU SemCom scenarios: when channel degradation induces severe signal distortion, the unconditional diffusion model may produce visually plausible yet semantically inaccurate images, for instance, reconstructing a “cat” as a “dog”, since both objects can occupy proximate regions within the pixel distribution manifold.

To overcome this challenge, CTD-Diff introduces a conditional diffusion reconstruction mechanism. Specifically, it extracts high-level semantic features from the aggregated semantic observation $\hat{X}_k$ using a semantic encoder and integrates these features as conditional inputs into the denoising network. This approach constrains the generation process to a manifold subspace that aligns with the original semantic content. Consequently, this design not only enhances the semantic fidelity of the reconstructed images but also improves the model’s robustness against channel noise and fading effects.

The fundamental components of the conditional diffusion network comprise the semantic encoder and the conditional denoising network. The semantic encoder utilizes a pre-trained ResNet50 as its backbone, capitalizing on its general feature extraction capabilities acquired from extensive visual tasks. Specifically, the convolutional layers of ResNet50 are frozen to retain the pre-trained knowledge, and its output is subjected to global average pooling to produce a feature vector. This vector is subsequently projected into a $d$-dimensional semantic embedding space through a trainable linear projection layer. Formally, we define the clean semantic embedding derived from original features as $z_k=f_{\mathrm{sem}}(X_k;\phi)\in\mathbb{R}^d$, and the estimated embedding derived from the noisy observation as $\hat{z}_k=f_{\mathrm{sem}}(\hat{X}_k;\phi)$. The semantic embedding encapsulates global content descriptors of the image, encompassing high-level attributes such as object categories, texture patterns, and color distributions. These features demonstrate a degree of robustness to noise and distortion. Therefore, even when the input is the noisy observation $\hat{X}_{k}$, the extracted embedding $\hat{z}_{k}$ maintains high semantic consistency with the original image $X_k$.

\begin{figure}[ht]
	\vspace{-2pt}
	\centering
	\includegraphics[scale=0.8]{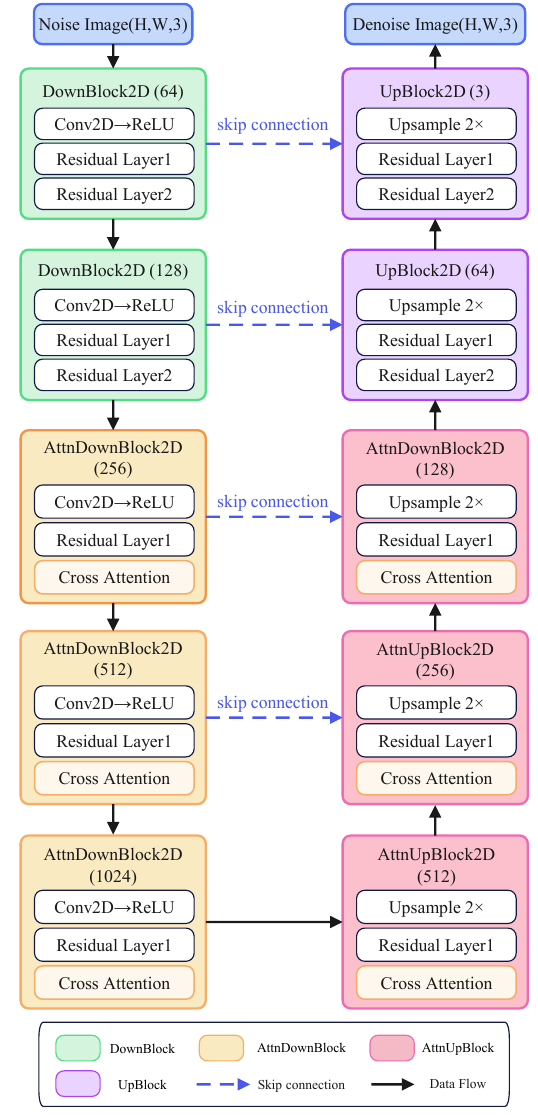}
	\caption{Architecture of the proposed conditional diffusion network in CTD-Diff.}
	\label{fig:Diffusion network}
	\vspace{-12pt}
\end{figure}

The conditional denoising network is designed based on the UNet architecture and consists of five downsampling stages followed by five upsampling stages, as illustrated in Fig.~\ref{fig:Diffusion network}. The initial two downsampling modules are standard \textit{DownBlock2D} units, primarily responsible for extracting low-level visual features such as edges and textures. The subsequent three downsampling modules utilize \textit{AttnDownBlock2D}, which integrate cross-attention mechanisms after the convolutional layers. This design enables the network to selectively emphasize task-relevant feature dimensions informed by the conditional embeddings. Consequently, the network can dynamically modulate its dependence on semantic information across different spatial locations, thereby enhancing global semantic consistency while preserving fine-grained details.

The decoder mirrors this architecture symmetrically. The first three upsampling modules are \textit{AttnUpBlock2D}, which continue to employ semantic embeddings to guide feature reconstruction. The final two upsampling modules are \textit{UpBlock2D}, tasked with restoring pixel-level details. Multiscale features are transmitted between the encoder and decoder through skip connections, ensuring the preservation of spatial information throughout the deep semantic processing stages.

With the introduction of semantic conditions, the reverse transfer kernel is formulated as
\begin{equation}
	p_\theta(X_{k,t-1}\mid X_{k,t},\hat{z}_k)=\mathcal{N}\left(\mu_\theta(X_{k,t},t,\hat{z}_k),\sigma_t^2I\right),
\end{equation}
\noindent where the mean function depends on the conditional embedding
\begin{equation}
	\mu_\theta(X_{k,t},t,\hat{z}_k)=\frac{1}{\sqrt{1-\beta_t}}\left(X_{k,t}-\frac{\beta_t}{\sqrt{1-\bar{\alpha}_t}}\epsilon_\theta(X_{k,t},t,\hat{z}_k)\right).
\end{equation}
\noindent Here, $\epsilon_\theta(X_{k,t},t,\hat{z}_k)$ represents the conditional noise prediction network, which injects the semantic embeddings $\hat{z}_{k}$ into the intermediate feature space via cross-attention layers. The training objective is reformulated as a conditional noise prediction loss
\begin{equation}
	\mathcal{L}_{\mathrm{cond}}=\mathbb{E}_{X_k,\epsilon,t}\left[\|\epsilon-\epsilon_\theta(X_{k,t},t,z_k)\|_2^2\right].
\end{equation}
\noindent It is important to note the distinction in the conditioning variable: during training, we employ the true semantic embeddings $z_{k}$ extracted from clean images to stabilize optimization; whereas during inference, we rely on the approximate embeddings $\hat{z}_{k}$ derived from the noisy observations $\hat{X}_k$. This discrepancy does not present substantial practical challenges, as semantic embeddings demonstrate robustness to noise. Furthermore, the network is exposed to samples with varying noise levels during training, enabling it to adapt effectively to input uncertainty.

The conditional diffusion model estimates the conditional distribution $p_{\theta}(X_{k}\mid\hat{z}_{k})$ rather than the marginal distribution $p_\theta(X_{k})$. This approach constrains the generative process to a subspace aligned with the semantic embedding $\hat{z}_{k}$, thereby substantially reducing the dimensionality of the search space. As a result, the model achieves improved accuracy and consistency in reconstruction. Moreover, under low SNR conditions, the conditional diffusion model preserves high semantic consistency, as the semantic embeddings serve as a global guide that directs the generation process appropriately, mitigating semantic drift issues commonly observed in unconditional models.

\subsection{Hybrid-Noise Training Strategy}
\label{sec:Hybrid Noise Training}

The efficacy of diffusion models is critically contingent upon the congruence between the noise distributions encountered during training and those present in testing data. Standard diffusion models utilize solely isotropic Gaussian noise in the training phase, wherein the noise introduced in the forward process adheres to a simple distribution, specifically $\mathcal{N}(0,(1-\bar{\alpha}_{t})I)$. In contrast, wireless SemCom systems experience receiver-side distortions that are substantially more intricate than Gaussian noise. These distortions include multiplicative effects arising from channel fading, correlated noise due to multipath propagation, noise amplification resulting from relaying, and additional non-Gaussian noise components. Consequently, when diffusion models trained exclusively on Gaussian noise are deployed under such realistic channel conditions, significant performance deterioration ensues, attributable to a mismatch between the denoising mechanisms learned by the network and the actual noise characteristics.

To address this challenge, CTD-Diff introduces a hybrid noise training methodology. By concurrently incorporating authentic channel noise and synthetic diffusion noise during training, the denoising network acquires a more generalized joint denoising operator, thereby sustaining robust performance across a variety of channel environments.

The fundamental concept underlying mixed noise training involves constructing a composite noise source that statistically encompasses the joint distribution of real channel noise and diffusion noise. Specifically, for each training sample $X_k$ and time step $t$, we first simulate the channel transmission process to generate the aggregated channel noise vector. This procedure entails randomly sampling Rayleigh fading coefficients, calculating the effective noise for each link after pre-equalization, and aggregating them. The resulting channel noise is then normalized to unit variance, denoted as $\boldsymbol{\epsilon}_{\mathrm{ch}}$.

Concurrently, a standard diffusion noise $\boldsymbol{\epsilon}_{\mathrm{df}}\sim\mathcal{N}(0,I)$ is independently sampled. The mixed noise is then defined as a convex combination of these two components
\begin{equation}
	\boldsymbol{\epsilon}_{\mathrm{hyb},t}=\lambda_t\boldsymbol{\epsilon}_{\mathrm{ch}}+\sqrt{1-\lambda_t^2}\boldsymbol{\epsilon}_{\mathrm{df}},\quad0\leq\lambda_t\leq1,
\end{equation}
\noindent where the mixing coefficient $\lambda_{t}$ regulates the relative contributions of channel noise and diffusion noise. Incorporating this mixed noise into the forward diffusion equation results in the formulation used to generate training samples
\begin{equation}
	X_{k,t}=\sqrt{\bar{\alpha}_{t}}X_k+\sqrt{1-\bar{\alpha}_{t}}\boldsymbol{\epsilon}_{\mathrm{hyb},t}.
\end{equation}
\noindent Note that unlike inference which starts from the received signal, the training forward process originates from the clean data $X_k$. The resulting sample $X_{k,t}$ encompasses both the statistical properties of actual channel distortion and the stochastic nature of the diffusion process. When training the denoising network on these samples, the objective is to predict the hybrid noise $\boldsymbol{\epsilon}_{\mathrm{hyb},t}$, with the loss function defined as follows
\begin{equation}
	\mathcal{L}_{\mathrm{hyb}}=\mathbb{E}_{X_k,\boldsymbol{\epsilon}_{\mathrm{hyb}},t}\left[\|\boldsymbol{\epsilon}_{\mathrm{hyb},t}-\epsilon_\theta(X_{k,t},t,z_k)\|_2^2\right].
\end{equation}
\noindent Mixed noise training corresponds to maximizing the likelihood of a composite generative model characterized by its noise distribution
\begin{equation}
	p_{\mathrm{hyb}}(\epsilon)=\lambda_t^2p_{\mathrm{ch}}(\epsilon)+(1-\lambda_t^2)p_{\mathrm{df}}(\epsilon),
\end{equation}
\noindent where $p_{\mathrm{ch}}$ denotes the empirical distribution of the channel noise, and $p_{\mathrm{df}}$ represents the standard Gaussian prior. The solution that minimizes $\mathcal{L}_{\mathrm{hyb}}$ corresponds to the minimum variance unbiased estimator:
\begin{equation}
	\epsilon_\theta^*(X_{k,t},t)=\lambda_t^2\mathbb{E}_{p_{\mathrm{ch}}}[\epsilon\mid X_{k,t},t]+(1-\lambda_t^2)\mathbb{E}_{p_{\mathrm{df}}}[\epsilon\mid X_{k,t},t].
\end{equation}
The denoising strategy learned by the network constitutes a weighted average of the inverse mapping of channel distortion and the diffusion reversal process, with the weights dynamically modulated by $\lambda_{t}$. The scheduling of $\lambda_t$ is designed to balance the effects of channel-induced noise and diffusion noise throughout the denoising trajectory. Deriving an optimal scheduling strategy is intractable due to the joint noise distribution, resulting from both unknown and dynamically varying wireless channels and semantic compression. Consequently, a linear decay schedule is employed as a practical approximation. This approach has been widely adopted in diffusion-based models to ensure stable training and effective noise annealing.
\cite{2020denoising, NEURIPS2024}

The selection of the hybrid coefficient $\lambda_{t}$ significantly impacts training effectiveness. If $\lambda_{t}$ is excessively large, the network tends to overfit the channel statistics, diminishing generalization. Conversely, if $\lambda_{t}$ is too small, the network reduces to a conventional diffusion model, inadequate for processing real-world channel noise. To address this, a time-dependent scheduling strategy is employed: during the initial diffusion phase, $\lambda_{t}$ is assigned a high value to simulate channel dominance; as $t$ increases, $\lambda_{t}$ is decreased. A straightforward effective scheduling is linear decay $\lambda_{t} = \lambda_{0}(1 - t/T)$.

Training with mixed noise additionally imparts a regularization effect. Because each training batch comprises a mixture of various channel realizations, the network develops a generalized denoising representation invariant to specific noise types. This heterogeneity improves resilience to input perturbations. The overall procedure is outlined in Algorithm \ref{alg:tdcdiff_training}.

\begin{algorithm}[t]
	\caption{Hybrid-Noise Conditional Diffusion Training of CTD-Diff}
	\label{alg:tdcdiff_training}
	\hspace*{0.02in}{\bf Input:}
	Dataset $\mathcal{D}$, diffusion steps $T$, schedule $\{\beta_t\}$, channel model $\mathcal{H}$\\
	\hspace*{0.02in}{\bf Output:}
	Trained denoiser $\boldsymbol{\epsilon}_\theta$
	\begin{algorithmic}[1]
		\State Initialize $\theta$
		\While{training not converged}
		\State Sample user $k$ and clean image $X_k\sim\mathcal{D}$
		\State Extract \textbf{clean} semantic condition $z_k=f_{\mathrm{sem}}(X_k;\phi)$
		\State Sample diffusion step $t\sim\mathrm{Uniform}(1,T)$
		
		\State Sample channel coefficients $h \sim \mathcal{H}$
		\State Generate normalized channel noise $\boldsymbol{\epsilon}_{\mathrm{ch}}$
		\State Sample Gaussian noise $\boldsymbol{\epsilon}_{\mathrm{df}}\sim\mathcal{N}(0,I)$
		\State Compute $\boldsymbol{\epsilon}_{\mathrm{hyb}} = \lambda_t \boldsymbol{\epsilon}_{\mathrm{ch}} + \sqrt{1-\lambda_t^2}\,\boldsymbol{\epsilon}_{\mathrm{df}}$
		
		\State Forward Process from Clean Data:
		\State $X_{k,t} = \sqrt{\bar{\alpha}_t}\,X_k + \sqrt{1-\bar{\alpha}_t}\,\boldsymbol{\epsilon}_{\mathrm{hyb}}$
		
		\State Predict noise $\hat{\boldsymbol{\epsilon}} = \boldsymbol{\epsilon}_\theta(X_{k,t}, t, z_k)$
		\State Update $\theta$ by minimizing $\|\boldsymbol{\epsilon}_{\mathrm{hyb}}-\hat{\boldsymbol{\epsilon}}\|_2^2$
		\EndWhile
		\Return $\boldsymbol{\epsilon}_\theta$
	\end{algorithmic}
\end{algorithm}

In summary, the CTD-Diff framework facilitates multi-user collaborative SemCom through three key innovations: (i) at the overall framework level, it integrates collaborative transmission into the diffusion generation process to establish a unified cross-layer model. (ii) at the network architecture level, it incorporates directional generation with semantic conditional constraints to improve reconstruction fidelity. (iii) at the training strategy level, it employs hybrid noise to bridge the gap between theoretical constructs and practical applications, thereby enhancing generalization capabilities. These three innovations are mutually reinforcing and collectively constitute the technical foundation of CTD-Diff, enabling high-quality semantic information reconstruction in complex MU fading channel environments.

\section{Numerical Experiments}
\label{sec:Experiments}
In this section, we evaluate the proposed CTD-Diff framework through comprehensive numerical simulations conducted under various channel conditions and datasets. The objective is to demonstrate that the proposed CTD-Diff system facilitates robust image reconstruction in MU scenarios, surpassing the performance of conventional source–channel coding and SemCom methods. We first introduce the datasets and baseline methods, followed by a description of the experimental configurations. Subsequently, we provide both quantitative and visual comparison results, along with ablation studies that elucidate the contributions of the key components of CTD-Diff.

\vspace{-8pt}
\subsection{Experiments Setting}

\subsubsection{Dataset}

In our experiments, the proposed CTD-Diff framework is evaluated on three image datasets of increasing complexity to ensure a comprehensive assessment of its reconstruction capabilities. The CIFAR100 dataset comprises 60,000 natural images of size $32×32$ across 100 diverse object categories. STL-10 contains images of higher resolution $96×96$ from 10 labeled classes. This dataset presents more complex spatial structures and greater background variability than CIFAR100. ImageNet-256, a subset of the large-scale ImageNet dataset with all images resized to $256×256$ pixels, constitutes a challenging benchmark characterized by substantial intra-class variation and fine-grained visual details. For all datasets, images are normalized to $[0,1]$ and randomly partitioned into training, validation, and testing sets. Prior to input into the semantic encoder, all images are resized to the target resolution $(H, W)$. The combined use of these datasets allows for evaluation of CTD-Diff across varying levels of semantic abstraction and texture complexity, thereby ensuring that performance conclusions generalize across simple, medium, and high-resolution visual domains.

\subsubsection{Parameters Setting}

All experiments are implemented in PyTorch 2.6.0 with CUDA 12.4 on a Linux server equipped with an AMD EPYC 7742 CPU and an NVIDIA V100-32GB GPU. The proposed CTD-Diff model consists of a frozen ResNet-50 semantic encoder and a conditional diffusion model. During training, we use AdamW as the optimizer with a learning rate of $1\times10^{-4}$ and cosine annealing scheduling. Mixed precision training  is enabled to accelerate computation and reduce memory cost. The diffusion process adopts a DDPM scheduler with 1000 timesteps during training. The training batch size is set to 20,  and we train the model for 50,000 epochs. During inference, each user transmits in a TDMA manner, and cooperative forwarding adopts the proposed pre-equalization strategy with direct signal aggregation at the receiver to leverage cooperative benefits.

\subsubsection{Baseline Schemes}

For comparison, we consider the typical communication method, the deep learning coding scheme and the MU SemCom method in a MU wireless setting. We compare CTD-Diff against three representative categories of MU transmission schemes. All baselines use the same TDMA structure for fair comparison.

\begin{itemize}
	\item JPEG compression combined with LDPC channel coding, where each user independently transmits in its TDMA slot without cooperation. This baseline represents the conventional digital communication pipeline.
	\item MU MVJSCC model \cite{MVJSCC}, where each user has an independent encoder producing continuous complex channel symbols, and the BS jointly decodes all users’ features. No overhearing or relaying is enabled. This baseline is a strong end-to-end semantic transmission method.
	\item Semantic Importance-Aware Communication (SIAC) framework \cite{Semanticimportance}, which estimates semantic importance using the encoder feature maps and allocates channel protection accordingly through unequal error protection and importance-aware power distribution. Each user still transmits only in its own TDMA slot, without cooperation. This baseline represents importance-driven MU SemCom under the same resource budget.
	\item Channel Denoising Diffusion Model (CDDM) baseline, where each user independently transmits its deep JSCC encoded features in its TDMA slot. At the BS, a diffusion model serves as a channel denoising post-processor. The denoised features are then decoded by the joint decoder.
\end{itemize} 

The proposed scheme differs from all baselines since users overhear in other users’ TDMA slots and forward the received waveform, enabling the BS to perform semantic feature aggregation and conditional diffusion based semantic reconstruction.

Reconstruction quality is assessed using the PSNR and multi–scale structural similarity (MS–SSIM). For an original image $\mathbf{I}_k$ and reconstructed image $\hat{\mathbf{I}}_k$, PSNR is defined as 
\begin{equation}
	\mathrm{MSE}
	= \frac{1}{C H W}
	\sum_{c=1}^{C}
	\sum_{i=1}^{H} \sum_{j=1}^{W}
	\left( I_{k,c,i,j} - \hat{I}_{k,c,i,j} \right)^2
\end{equation}

\begin{equation}
	\mathrm{PSNR}
	= 10 \log_{10} \left( \frac{MAX^2}{\mathrm{MSE}_k} \right)
\end{equation}

where $C=3$ indexes the RGB color channels, and $MAX$ denotes the maximum pixel value. MS–SSIM evaluates perceptual structural fidelity across multiple scales and is defined as
\begin{equation}
	\begin{aligned}
		\mathrm{MS\text{-}SSIM}(\mathbf{I}_k,\hat{\mathbf{I}}_k)
		&= \left[ l_M(\mathbf{I}_k,\hat{\mathbf{I}}_k) \right]^{\alpha_M} \\
		&\quad \times \prod_{m=1}^{M}
		\left[ c_m(\mathbf{I}_k,\hat{\mathbf{I}}_k) \right]^{\beta_m}
		\left[ s_m(\mathbf{I}_k,\hat{\mathbf{I}}_k) \right]^{\gamma_m}.
	\end{aligned}
\end{equation}
where $M$ denotes the number of scales, $l_m$, $c_m$, and $s_m$ represent luminance, contrast and structure components at scale $m$. The parameters $\alpha_{M}$, $\beta_{m}$, and $\gamma_{m}$ are non–negative weighting exponents that control the relative importance of the luminance, contrast, and structure components at each scale, and are set following standard practice. Higher PSNR and MS–SSIM values indicate better reconstruction quality and greater robustness under MU fading channels.

\vspace{-6pt}
\subsection{Performance Analysis}
\vspace{-2pt}

\begin{figure*}[ht]
	\vspace{-12pt}
	\centering
	\includegraphics[scale=0.21]{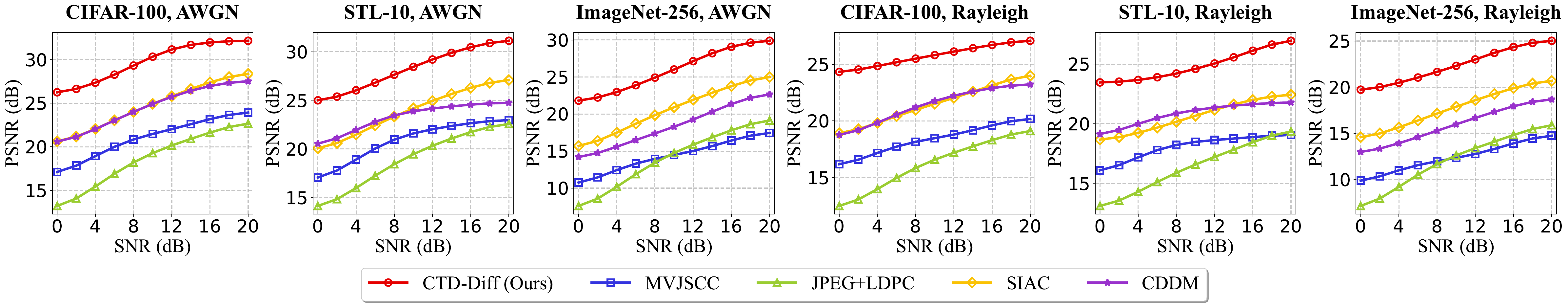}
	\caption{Comparison of the PSNR performance in different datasets with AWGN and Rayleigh fading.}
	\label{fig:psnr_comparison}
	\vspace{-4pt}
\end{figure*}

\begin{figure*}[ht]
	\centering
	\vspace{-2pt}
	\includegraphics[scale=0.21]{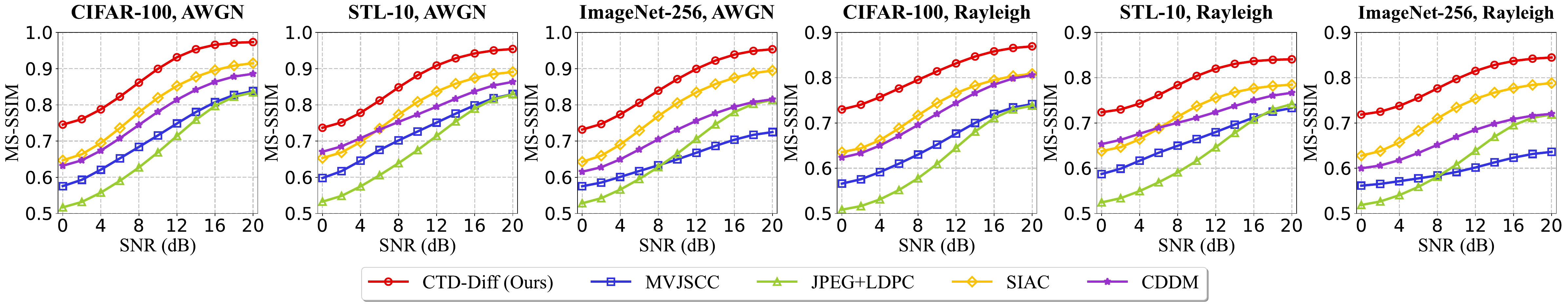}
	\caption{Comparison of the MS-SSIM performance in different datasets with AWGN and Rayleigh fading.}
	\label{fig:ssim_comparison}
	\vspace{-12pt}
\end{figure*}

Figs. \ref{fig:psnr_comparison} and \ref{fig:ssim_comparison} illustrate the PSNR and MS-SSIM performance of the proposed CTD-Diff framework in comparison with three representative MU baselines: JPEG+LDPC, MU MVJSCC model~\cite{MVJSCC}, and SIAC, under Additive white Gaussian noise (AWGN) and Rayleigh fading channels, with the number of users fixed at $K=20$. The experiments utilized the CIFAR-100, STL-10, and ImageNet-256 datasets, representing low, medium, and high-resolution image domains. Across all datasets and channel conditions, CTD-Diff consistently outperforms the baselines, achieving superior reconstruction fidelity and demonstrating notable advantages in robustness, semantic preservation, and scalability in MU scenarios.

\subsubsection{PSNR Performance Analysis}

On AWGN channels, CTD-Diff demonstrates a significant PSNR improvement compared to all baseline methods. For the CIFAR-100 and STL-10 datasets, CTD-Diff outperforms MVJSCC by approximately 2 to 4 dB across the entire SNR spectrum, with this performance gap increasing to over 5 dB on the ImageNet-256 dataset. This enhanced advantage is more evident at higher image resolutions, indicating CTD-Diff’s capacity to exploit conditional diffusion priors to recover fine-grained textures and high-frequency semantic details that baseline models tend to oversmooth or lose due to quantization and bit errors. Although SIAC achieves better results than JPEG combined with LDPC coding through importance-aware error protection, it still underperforms relative to CTD-Diff, as the digital transmission pipeline remains vulnerable to residual channel noise and semantic inconsistencies in complex image content.

Under Rayleigh fading conditions, the advantages of CTD-Diff become increasingly pronounced. Severe amplitude fluctuations substantially degrade the performance of baseline methods, particularly JPEG combined with LDPC coding, whose PSNR deteriorates markedly at low SNR levels due to burst errors. While MVJSCC demonstrates moderate robustness, it encounters difficulties in compensating for deep fades in the absence of cooperative mechanisms. In contrast, CTD-Diff consistently maintains stable PSNR across all SNR regimes, achieving improvements of up to 6–8 dB on the ImageNet-256 dataset. This enhanced robustness can be attributed to two principal design features: (i) cooperative overhearing and amplify-and-forward relaying, which enhance the effective received quality through coherent signal aggregation, and (ii) hybrid-noise trained conditional diffusion, which effectively models realistic channel distortions and reconstructs missing structural information by leveraging semantic priors.

\subsubsection{MS-SSIM Performance Analysis}

The MS-SSIM results corroborate the PSNR trends from a perceptual standpoint. Across all datasets, CTD-Diff consistently attains MS-SSIM values exceeding 0.9 at moderate to high SNR levels, demonstrating robust preservation of object contours, color fidelity, and multi-scale structural characteristics. MVJSCC achieves competitive MS-SSIM performance at high SNR. However, its performance deteriorates significantly under Rayleigh fading conditions due to unstable joint decoding of MU features. SIAC shows enhanced structural preservation relative to JPEG+LDPC, yet its performance remains limited by discrete semantic-importance allocation and the absence of cooperation.

In contrast, CTD-Diff leverages its conditional diffusion mechanism, which incorporates semantic embeddings derived from aggregated observations. This approach facilitates perceptually coherent reconstructions despite declines in pixel-wise SNR, resulting in the highest MS-SSIM across all datasets. The advantage is especially pronounced on ImageNet-256, where the increased complexity of images accentuates the limitations inherent in purely feature-based or digital baseline methods.

A fundamental characteristic of CTD-Diff is its implementation of MU overhearing and relay forwarding within TDMA time slots. While all baseline methods operate under an identical total resource budget and utilize the same number of TDMA slots, they treat users independently, which constrains the effective SNR and precludes improvements. In contrast, CTD-Diff reconfigures TDMA into a cooperative framework wherein each user's transmission is monitored by neighboring users, relayed to the BS, and coherently combined via direct aggregation. This approach yields improvements that, when integrated with semantic-conditioned diffusion denoising, substantially enhance reconstruction stability which is a benefit consistently demonstrated across all performance curves.

Across PSNR and MS-SSIM, under both AWGN and Rayleigh channels, and across datasets of varying complexity, CTD-Diff demonstrates several consistent advantages. First, it achieves the highest fidelity according to both pixel-wise and perceptual metrics, thereby validating the effectiveness of diffusion-based semantic reconstruction. Second, it exhibits superior robustness in fading and low-SNR regimes, attributable to cooperative transmission and hybrid-noise training. Third, it shows enhanced scalability when applied to complex, high-resolution datasets, where the incorporation of semantic priors and multi-scale denoising is essential. Fourth, it consistently outperforms both traditional digital pipelines and end-to-end neural network baselines.

The findings collectively confirm that the proposed CTD-Diff framework effectively utilizes TDMA cooperation, conditional diffusion modeling, and hybrid-noise robustness to attain significant performance in MU SemCom systems.

\vspace{-8pt}
\subsection{Ablation Studies}

\subsubsection{Ablation Studies on Cooperative Mechanism}

To isolate the contributions of cooperation and the associated performance benefits within the proposed CTD-Diff framework, we conducted a series of ablation studies under controlled conditions. The experiments compared the following scenarios: (i) CTD-Diff with the cooperation mechanism, wherein idle users overhear transmissions and forward relayed signals to facilitate aggregation prior to diffusion reconstruction; and (ii) CTD-Diff without the cooperation mechanism, in which each user transmits solely during its assigned TDMA slot, with no overhearing or relaying, resulting in the BS receiving only a single direct copy.

\begin{figure}[tb]
	\vspace{-2pt}
	\centering
	\includegraphics[scale=0.338]{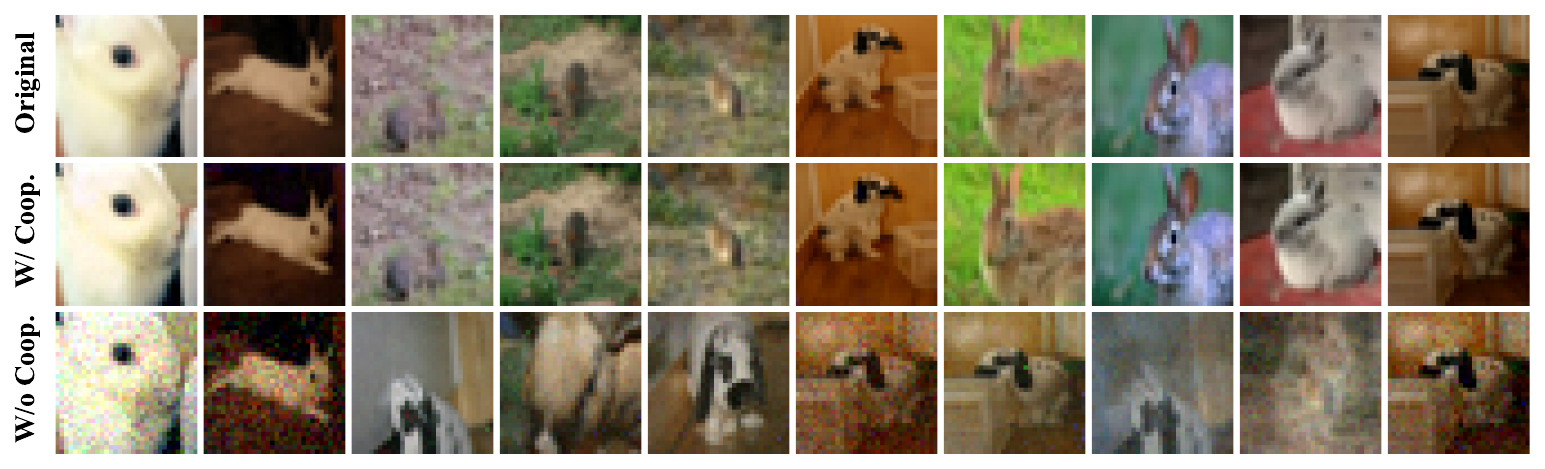}
	\caption{Reconstruction Comparison With vs. Without Cooperation on the CIFAR-100 dataset for 10 users at 10 dB AWGN channel.}
	\vspace{-6pt}
	\label{fig:withcoop}
	\vspace{-12pt}
\end{figure}

These experiments demonstrate the direct benefits conferred by cooperative processing prior to semantic diffusion reconstruction. The visual examples presented in Fig. \ref{fig:withcoop} demonstrate that incorporating cooperation substantially enhances perceptual quality. In the absence of cooperation, reconstructed images display prominent distortions, including blurred edges, color inconsistencies, and structural inaccuracies, particularly in textured areas. Conversely, cooperative forwarding increases the effective SNR at the BS, leading to sharper local textures, more precise color reproduction, and improved preservation of semantic structures. This qualitative enhancement underscores the advantage of initiating the diffusion process with an aggregation enhanced observation rather than relying on a single noisy copy.

\begin{figure}[ht]
	\vspace{-12pt}
	\centering
	\includegraphics[scale=0.32]{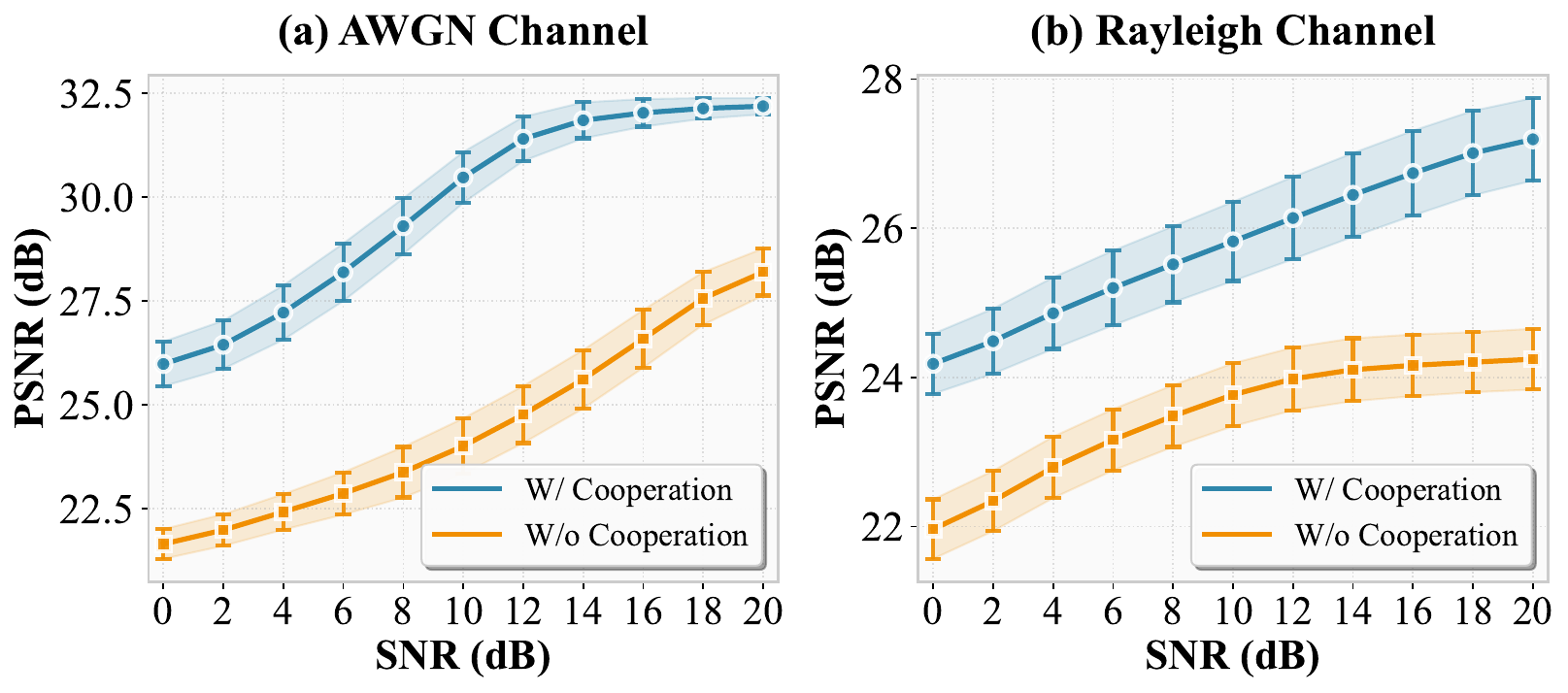}
	\caption{PSNR Comparison Under Cooperation vs. No Cooperation on the CIFAR-100 dataset for 20 users at 10 dB AWGN channel.}
	\label{fig:psnrmultiuser}
	\vspace{-6pt}
\end{figure}

\begin{figure}[ht]
	\vspace{-2pt}
	\centering
	\includegraphics[scale=0.32]{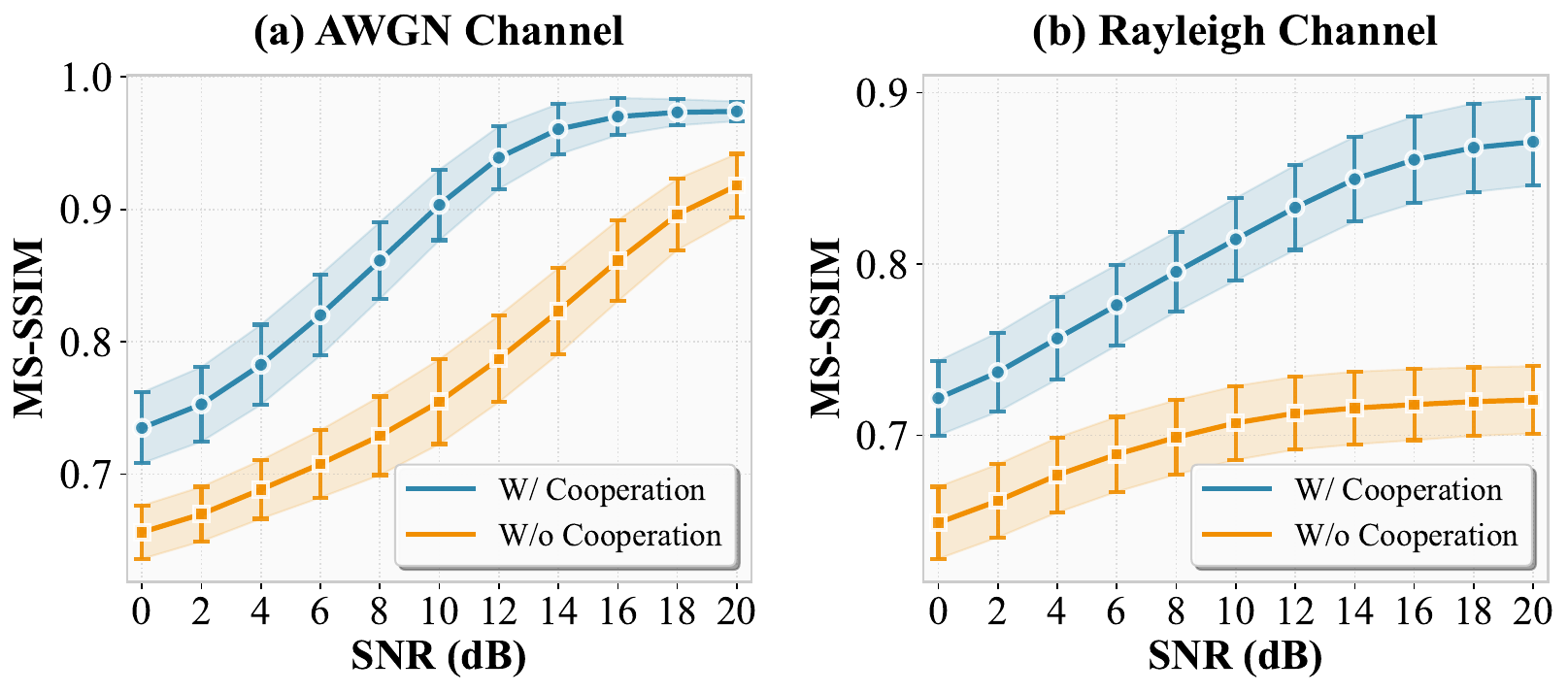}
	\caption{MS-SSIM Comparison Under Cooperation vs. No Cooperation on the CIFAR-100 dataset for 20 users at 10 dB AWGN channel.}
	\label{fig:ssimmultiuser}
	\vspace{-12pt}
\end{figure}

The curves with standard deviation bars presented in Figs.~\ref{fig:psnrmultiuser} and \ref{fig:ssimmultiuser} illustrate consistent and statistically significant performance improvements when cooperation is enabled. Across all SNR levels, cooperative CTD-Diff achieves an enhancement of 3.8 to 4.2 dB in PSNR. This improvement is most pronounced at low SNR values, where aggregation provides substantial improvement. Additionally, the reduction in error bar size with cooperation indicates decreased performance variance, suggesting enhanced reliability among users experiencing varying channel fading conditions.

MS-SSIM increases by 10\% to 16\%, indicating a more accurate restoration of structural information. High-level semantic features, including object identity, geometry, and shape, are significantly better preserved when cooperation is employed. This observation is consistent with the hypothesis that semantic embeddings derived from a cleaner aggregated semantics offer a more stable conditioning vector for diffusion-based reconstruction. Collectively, these results demonstrate that cooperation not only enhances average performance but also stabilizes reconstruction quality across different users.

\subsubsection{Ablation Studies on User Quantity}

\begin{figure}[htp]
	\vspace{-2pt}
	\centering
	\includegraphics[scale=0.37]{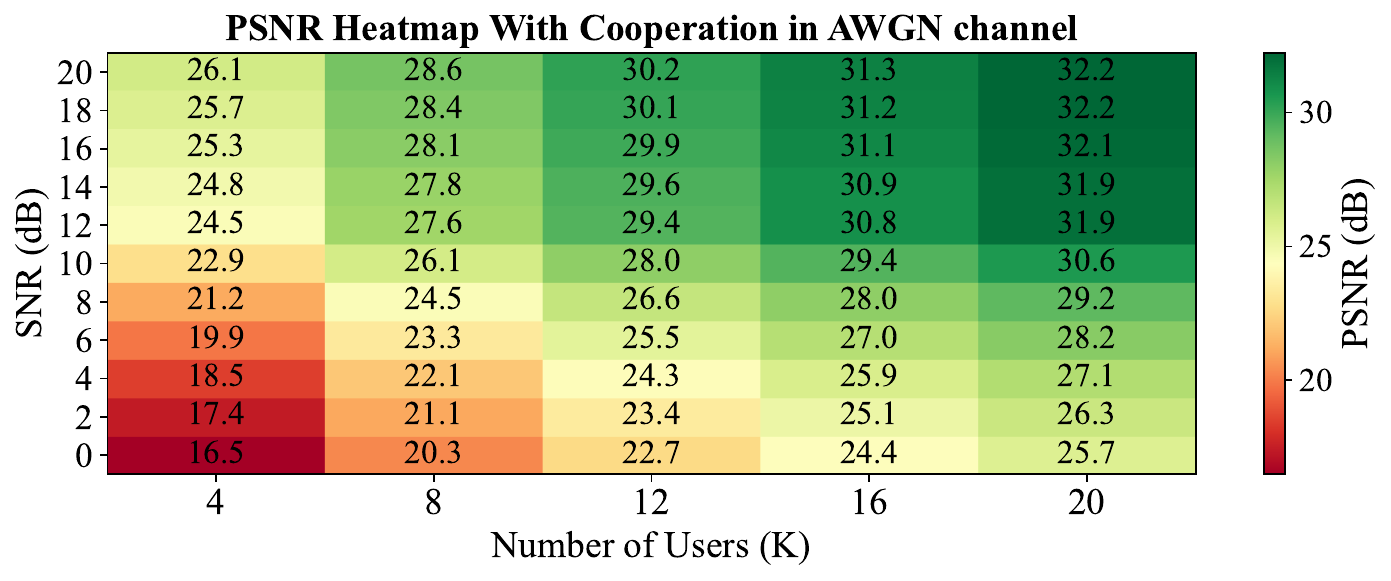}
	\caption{PSNR Heatmap Across User Quantity in AWGN Channel.}
	\vspace{-6pt}
	\label{fig:Heatmap awgn}
	\vspace{-8pt}
\end{figure}

\begin{figure}[htp]
	\vspace{-8pt}
	\centering
	\includegraphics[scale=0.37]{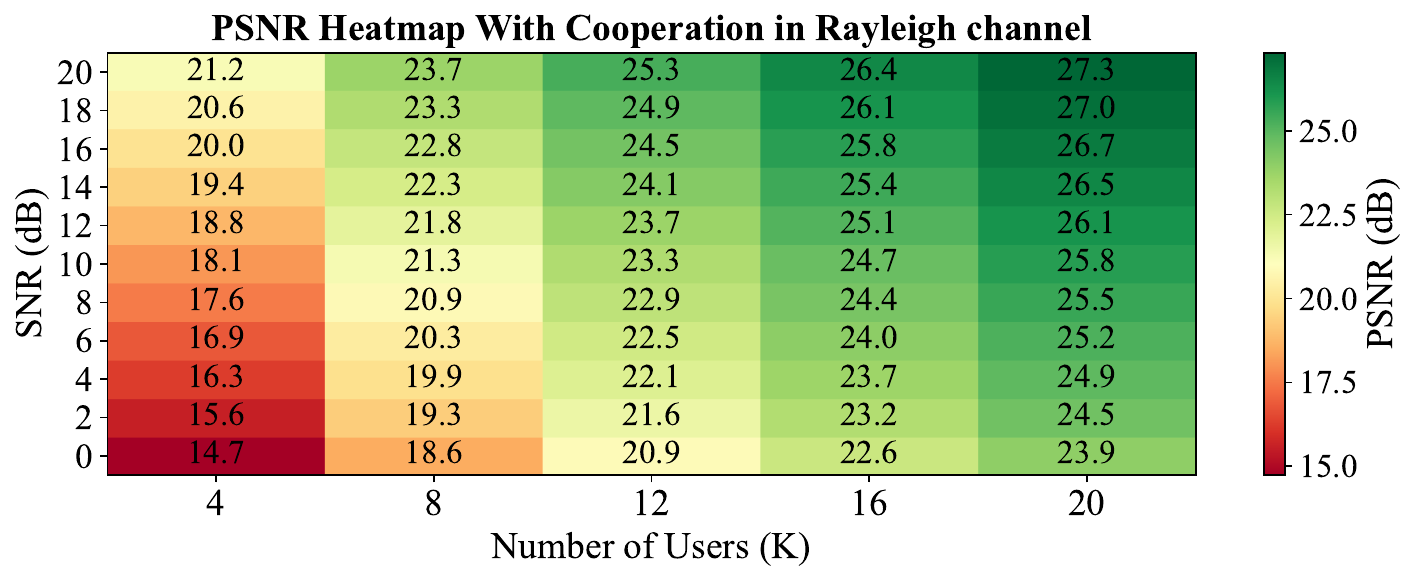}
	\caption{PSNR Heatmap Across User Quantity in Rayleigh Channel}
	\vspace{-6pt}
	\label{fig:Heatmap rayleigh}
	\vspace{-4pt}
\end{figure}

We further investigate the impact of cooperative benefits on the overall system performance by varying the number of users $K\in\{4,8,12,16,20\}.$ For each user number, cooperative relaying yields an increased number of overheard copies per active user, thereby enhancing the performance improvements. The results are presented in two heatmaps. As depicted in Fig.~\ref{fig:Heatmap awgn}, the PSNR improves monotonically with an increasing number of users. However, the improvements tend to saturate as $K$ approaches 20. When $K = 4$, the average PSNR remains relatively low due to limited overhearing opportunities. In contrast, at $K = 20$, the system achieves an improvement exceeding 4 dB in PSNR compared to $K = 4$, demonstrating that dense MU cooperation effectively aggregates clean energy from multiple relays.

As illustrated in Fig.~\ref{fig:Heatmap rayleigh}, Rayleigh fading induces more pronounced fluctuations, thereby enhancing the value of cooperative benefits. The cooperative scheme attains an approximate improvement of 5 to 6 dB when increasing the number of users from $K=4$ to $K=20$. This confirms that cooperation is particularly crucial in challenging fading environments where the direct communication link may experience significant attenuation. The heatmaps for both channels demonstrate a consistent pattern: an increase in the number of users leads to a greater number of overheard links, which in turn results in stronger cooperative benefits, improved diffusion initialization, and ultimately enhanced reconstruction quality. These results substantiate the design philosophy of CTD-Diff, emphasizing that MU cooperation is not merely an auxiliary feature but a fundamental mechanism that directly improves generative reconstruction.

The ablation studies indicate that cooperation is crucial for fully realizing the potential of diffusion-based semantic reconstruction. An increase in the number of users directly contributes to greater improvement, resulting in consistent improvements in PSNR and MS-SSIM. The interaction between cooperative physical-layer processing and semantic diffusion constitutes the cross-layer foundation underpinning the robustness of CTD-Diff. Collectively, these findings demonstrate that cooperative relaying is not merely an auxiliary physical-layer component but a fundamental mechanism that amplifies the efficacy of conditional diffusion models in MU SemCom.

\vspace{-2pt}
\section{Conclusion} 
\label{sec:conclusion}

In this paper, we propose CTD-Diff, a cooperative MU SemCom framework that integrates TDMA overhearing, aggregation-based semantics fusion, and conditional diffusion reconstruction. Additionally, we propose a hybrid-noise training strategy that unifies channel impairments and diffusion perturbations within a single generative modeling process. Experimental evaluations conducted on multiple datasets and under various channel conditions demonstrate that CTD-Diff consistently outperforms representative digital and semantic baselines across different SNR regimes. Notably, the cooperative transmission mechanism yields substantial improvements in low-SNR and fading environments, evidencing strong robustness and scalability in MU scenarios. These findings validate the effectiveness of the proposed cooperative generative SemCom framework and underscore its potential for reliable, high-fidelity image transmission in future MU wireless systems.

\ifCLASSOPTIONcaptionsoff
\newpage
\fi

\vspace{-4pt}
\bibliographystyle{IEEEtran}
\vspace{-2pt}
\bibliography{refer.bib}

\end{document}